\documentclass[aps, physrev, twocolumn, preprintnumbers,10pt]{revtex4-2}
\usepackage{graphicx}
\usepackage{dcolumn}
\usepackage{bm}
\usepackage{amsmath}
\usepackage{amssymb}
\usepackage{multirow}
\usepackage{subcaption}
\usepackage{caption}
\usepackage[colorlinks=true, allcolors=blue]{hyperref}

\begin{document}
\preprint{CTPU-PTC-24-42}
\title{\textbf{Probing $B^+ \to K^+$ semileptonic FCNC decay with new physics effects in the PQCD approach}}
\author{Chao-Qi Zhang$^1$}
    \email{E-mail: cqzhang@nnu.edu.cn}
\author{Jin Sun$^{2}$}
\email{E-mail: sunjin0810@ibs.re.kr(corresponding author)}
\author{Zhi-Peng Xing$^1$}
    \email{E-mail: zpxing@nnu.edu.cn}
\author{Rui-Lin Zhu$^1$}
    \email{E-mail: rlzhu@njnu.edu.cn} 
\affiliation{$^1$Department of Physics and Institute of Theoretical Physics, Nanjing Normal University, Nanjing 210023, Jiangsu, China}
\affiliation{$^2$Particle Theory and Cosmology Group, Center for Theoretical Physics of the Universe, Institute for Basic Science (IBS), Daejeon 34126, Korea }
\date{\today}

\begin{abstract}
Recently, the Belle II Collaboration reported the branching fraction $\mathcal{B}(B^+ \to K^+ \nu \bar{\nu})=(2.3\pm0.7)\times10^{-5}$ with a significance of $3.5\sigma$, which is $2.7\sigma$ above the Standard Model expectation.
Motivated by this measurement, we calculate this decay channel at the next-to-leading order and twist-three level using the perturbative QCD approach. 
By combining lattice QCD data with our results, we obtain form factors with improved reliability.
Using these form factors, we estimate the branching ratios of $B^+ \to K^+$ semileptonic flavor-changing neutral current decays, including  $B^+ \to K^+\nu\bar\nu$ and $B^+ \to K^+\ell^+\ell^-$. 
To address the anomalies in these two processes, we introduce a leptoquark model as a new physics scenario. Analyzing five possible types of leptoquarks, we successfully explain the latest experimental measurements and derive further constraints on the leptoquarks.
Furthermore, the single leptoquark model becomes invalid once the stringent constraints from process $B_s \to \mu^+\mu^-$ and $B_s$-$\bar{B_s}$ mixing is taken into account.
\end{abstract}

\maketitle

\section{Introduction}
The exclusive $B$ decay processes play a crucial role in probing the complex dynamics of both the quark and lepton sectors, serving not only as a powerful tool for precisely testing the Standard Model (SM) but also for searching for new physics (NP) effects.
Over the past 60 years, numerous important $B$ decay processes have been measured across various experimental facilities, with many results confirmed by SM predictions.
With the accumulation of more and more experimental data~\cite{CMS:2014xfa,LHCb:2014vgu,Belle:2016fev,LHCb:2017avl,Belle:2017oht,LHCb:2020lmf,LHCb:2021trn,ParticleDataGroup:2024cfk}, $B$ decay physics has firmly established itself as a leading field for precisely testing the SM and searching for NP.
 
In $B$ decay processes, semileptonic decays  characterized by the final state containing both hadron and a lepton pair are a major area of research. 
Since the lepton pair and hadron in the final state hardly interact with each other, this helps significantly reduce uncertainties in theoretical calculations. 
The decay amplitude in these processes is primarily determined by the hadronic matrix element, which can be expressed using form factors. 
Among these, flavor-changing neutral current (FCNC) processes, which have displayed significant flavor anomalies, serve as an ideal platform for precisely testing the SM and searching for NP.
The FCNC processes are naturally suppressed by the Glashow-Iliopoulos-Maiani mechanism and forbidden at tree level in the SM. 
The suppression from the SM makes these processes highly sensitive to contributions from NP effects.
Recently, with the substantial increase in experimental luminosity, the Belle II Collaboration has reported the first evidence for the $B^+ \to K^+ \nu \bar{\nu}$ decay with a significance of 3.5 standard deviations as~\cite{Belle-II:2023esi}
\begin{eqnarray}
    \mathcal{B}(B^+ \to K^+ \nu \bar{\nu})=(2.3\pm0.7)\times10^{-5}.
\end{eqnarray}
The latest experimental progress has greatly stimulated interest in theoretical research~\cite{Tian:2024ubt,Altmannshofer:2024kxb,Becirevic:2024iyi,Allwicher:2024ncl,Dash:2024crn,Wang:2024prt,Hati:2024ppg,Kim:2024tsm,Andersson:2024nam,Buras:2024ewl,Rosauro-Alcaraz:2024mvx,Karmakar:2024gla,Marzocca:2024hua,Bolton:2024egx,He:2024iju,Chen:2024cll,Hou:2024vyw,Li:2024thq,Gabrielli:2024wys,Loparco:2024olo,Chen:2024jlj,Ho:2024cwk,Fridell:2023ssf,McKeen:2023uzo,Altmannshofer:2023hkn,Datta:2023iln,Berezhnoy:2023rxx,He:2023bnk,Amhis:2023mpj,Abdughani:2023dlr,Allwicher:2023xba,Bause:2023mfe,Becirevic:2023aov,Athron:2023hmz}.

Over the past decade, $B$ decays have been extensively studied within the frameworks of perturbative QCD (PQCD)~\cite{Xiao:2013lia,Rui:2014tpa,Wang:2014qya,Liu:2016rqu,Rui:2016opu,Klein:2017xti,Liu:2017cwl,Wang:2017fen,Lu:2018cfc,Hua:2018pkq,Shen:2018abs,Su:2019vbu,Cheng:2020fcx,Bell:2020qus,Jin:2020jtu,Hua:2020usv,Beneke:2021jhp,Huber:2021cgk,Chai:2022ptk,Liu:2023kxr,Wang:2024xci}, lattice QCD (LQCD)~\cite{Bouchard:2013eph,Bouchard:2013mia,Bouchard:2014ypa,Flynn:2015mha,Bailey:2015dka,FermilabLattice:2015mwy,FermilabLattice:2019ikx,Colquhoun:2022atw,Martinelli:2022tte,Flynn:2023nhi}, and QCD sum rules (QCDSRs)~\cite{Wang:2015vgv,Shen:2016hyv,Rusov:2017chr,Gubernari:2018wyi,Zhou:2019jny,Shen:2021yhe}.
The PQCD approach, as a method for perturbative QCD analysis, serves as a robust tool for estimating the matrix elements of semileptonic decays of $B$ mesons in the low-momentum-transfer region~\cite{Li:1992nu,Li:1994iu,Kurimoto:2001zj}. 
It has demonstrated significant predictive power concerning $B$ decays, particularly regarding $CP$ violation~\cite{Li:2003yj,Li:2004ep,Ali:2007ff,Xiao:2011tx,Qi:2018lxy,Rui:2021kbn}. 
Regarding the FCNC process $B^+ \to K^+$, LQCD has achieved the most precise determination of the $B \to K$ form factor~\cite{Parrott:2022rgu}, consistent with earlier PQCD calculations that included twist-two and next-to-leading order (NLO) corrections~\cite{Wang:2012ab}. 
Nevertheless, using this highly accurate form factor, the theoretical prediction for the branching ratio of $B^+ \to K^+\nu\bar \nu$ differs from the experimental measurement by approximately 2.7 standard deviations~\cite{Belle-II:2023esi} as
\begin{eqnarray}
    \frac{\mathcal{B}(B^+ \to K^+ \nu \bar{\nu})|_{\mathrm{Exp}}}
    {\mathcal{B}(B^+ \to K^+ \nu \bar{\nu})|_{\mathrm{LQCD}}}=4.1\pm1.3.
\end{eqnarray}
The significant discrepancy between the theoretical predictions and experimental measurements necessitates considering higher-twist (twist-three) and higher-order (NLO) contributions within the PQCD framework, as well as potential NP effects in this process.

Among various NP models, leptoquarks (LQs) can introduce direct couplings between leptons and quarks~\cite{Pati:1973uk, Pati:1974yy}, thus bridging this fundamental connection between quark and lepton families. 
At low energies, LQs can induce two-lepton-two-quark interactions similar to those mediated by electroweak four-fermion vertices, making them a promising framework for explaining the previously mentioned discrepancies. 
LQs, as new color-triplet particles that can manifest as either scalar or vector bosons and carry both lepton and baryon numbers, have been extensively utilized to explain a range of experimental flavor anomalies~\cite{Bauer:2015knc,Becirevic:2018afm,Fajfer:2015ycq,Becirevic:2024pni,Chen:2023wpb}. These anomalies include $R_{K^{(*)}}$, $R_{D^{(*)}}$, and the muon's anomalous magnetic moment $(g-2)_\mu$ in scenarios where the LQ-lepton-quark couplings are not confined to a single generation. 
Additionally, searches for relevant collider signals have been conducted using various approaches~\cite{Buchmuller:1986zs,Dorsner:2016wpm}. 
LQs typically decay into either a charged lepton and a quark or a neutrino and a quark. 
At conventional particle-antiparticle colliders, LQs can be produced in pairs, setting a lower limit on their mass to approximately half the center-of-mass energy of the accelerator. 
However, heavier LQs can only participate in $t$-channel exchanges, resulting in a smaller cross section for the interactions they mediate. 
For instance, at the LHC, the leading-order (LO) processes for leptoquark pair production primarily involve gluon-gluon fusion and quark-antiquark annihilation. 

For the processes we have investigated, the FCNC $b\to s$ decays can occur via the exchange of a leptoquark as an intermediate state. 
In SM calculations, the uncertainties in our predictions can be significantly reduced by integrating LQCD data from high-momentum-transfer regions with PQCD calculations at low-momentum-transfers.
The improved reliability of SM predictions, combined with the significant discrepancy between experimental results and theoretical expectations, highlights the critical need to explore further constraints on leptoquark parameters, including their masses and couplings, within our study.


This paper is organized as follows: In Sec.~\ref{sec:framework}, we define the kinematics of the decay and give the necessary inputs for our calculations. 
In Sect.~\ref{sec:smcalc}, we present the framework for the theoretical calculations. 
In Sect.~\ref{sec:results}, we write down the observables of the process and present our numerical results and discussion in SM.
In Sec.~\ref{sec:npm}, we introduce the LQ model and place constraints on the NP parameters. 
In Sec.~\ref{sec:summary}, we make a brief summary. Some useful expressions are listed in the Appendix.

\section{Kinematics and Wave Functions}\label{sec:framework}
\begin{figure}[htbp]
    \centering
    \includegraphics[width=0.6\linewidth]{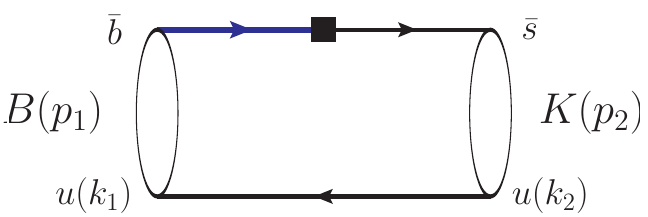}
    \caption{\label{fig:form} The Feynman diagrams for $B^+\to K^+$ transition which include the momenta of quark in initial and final meson states. 
    The black block represents vertices for weak interactions. The gluon propagators are omitted in this figure.}
\end{figure}

In our work, the theoretical calculations are given under the PQCD factorization framework. In the rest frame of the $B$ meson, the momentum of $B$ and $K$ can be defined as 
\begin{align}
    p_1=\frac{M_B}{\sqrt{2}}\left(1,1,\textbf{0}_{\mathrm{T}}\right),\quad
    p_2=\frac{M_B}{\sqrt{2}}\left(0,\eta,\textbf{0}_{\mathrm{T}}\right),
\end{align}
under the light cone coordinates, respectively. $M_B$ is the mass of the $B$ meson, and $\eta$ is the energy fraction carried by $K$. Here we define the transfer momentum $q=p_1-p_2$, then the energy fraction $\eta=1-q^2/M_B^2$. The momenta of light quarks in the $B$ and $K$ shown in Fig.~\ref{fig:form} are written as
\begin{align}
    k_1=\frac{M_B}{\sqrt{2}}\left(x_1,0,\textbf{k}_{1\mathrm{T}}\right),\quad
    k_2=\frac{M_B}{\sqrt{2}}\left(0,\eta x_2,\textbf{k}_{2\mathrm{T}}\right),
\end{align}
where $x_1$ and $x_2$ are parton momentum fractions. $\textbf{k}_{1\mathrm{T}}$ and $\textbf{k}_{2\mathrm{T}}$ are the transverse momentum. In this factorization approach, the scales that are involved in this processes can be divided as $m_b^2\gg m_b\Lambda_{\mathrm{QCD}}\gg\Lambda_{\mathrm{QCD}}^2$. The part with scale $m_b^2$ and $m_b\Lambda_{\mathrm{QCD}}$ can be perturbatively calculated, called the hard scattering kernel, and the rest parts are factorized as nonperturbative input called wave functions. This type of factorization may suffer end point singularities~\cite{Szczepaniak:1990dt,Burdman:1991xye,Beneke:2000wa}. A notable feature of PQCD is the preservation of transverse momentum $k_T$ to eliminate end point divergences~\cite{Ralston:2003mt}. In this strategy, the decay amplitude can be factorized into the convolution of the Wilson coefficient $C$, hard scattering kernel $H$, the hadronic wave functions $\Phi$, jet function $J_t$, and Sudakov factor $S$ as
\begin{eqnarray}
    {\cal A}=C\otimes H\otimes \Phi\otimes J_t\otimes S.
\end{eqnarray}
Here the jet function and Sudakov factor come from the threshold resummation and $k_T$ resummation, respectively~\cite{Li:1994cka}. Recently, the resummation schemes are systematically discussed in Refs.~\cite{Li:2012md,Li:2013xna}.
The hadronic wave functions, which are expressed by the light cone distribution amplitude (LCDA), can be extracted from experiments or other nonperturbative methods~\cite{Ball:2004ye,Ball:2006wn,Ball:2007rt,Bali:2017ude,RQCD:2019osh,Hua:2020gnw}. 

The $B$ meson distribution amplitude (DA) is defined via the matrix element as
\begin{eqnarray}
    &&\int\frac{d^4z}{(2\pi)^4}e^{ik\cdot z}\langle0|u_\alpha(z)\bar b_\beta(0)| B(p_1)\rangle=\notag\\
    &&\quad \frac{i}{\sqrt{2N_c}}\bigg((\rlap{/}p_1+M_B)\gamma_5\bigg[\phi_B(k_1)-\frac{\rlap{/} n-\rlap{/}v}{\sqrt{2}}\bar\phi_B(k_1)\bigg]\bigg)_{\beta\alpha},
\end{eqnarray}
where $n=(1,0,\bf 0_T)$ and $v=(0,1,\bf 0_T)$. $N_c=3$ is the number of colors. 
Then the wave function can be written as
\begin{eqnarray}
    \Phi_B=\frac{i}{\sqrt{2N_c}}(\rlap{/}p_1+M_B)\gamma_5\phi_B(k_1).
\end{eqnarray}
For the $\phi_B$, we adopt~\cite{Kurimoto:2001zj}
\begin{align}
    \phi_B(x,b)=N_Bx^2(1-x)^2\exp\left[-\frac{x^2M_B^2}{2\omega^2_B}
    -\frac{\omega^2_Bb^2}{2}\right],
\end{align}
where $\omega_B=0.40\pm0.04$ GeV is the shape parameter and $b$ is the parameter conjugate to $\textbf{k}_{\mathrm{T}}$. The normalization constant $N_B$ satisfied
\begin{align}
    \int_0^1\phi_B(x,b=0)dx=\frac{f_B}{2\sqrt{2N_c}},
\end{align}
with decay constant $f_B$. The light meson LCDAs are also defined through the matrix elements as
\begin{eqnarray}
    &&\langle K(p_2)|u_{\beta}(z)\bar s_{\alpha}(0)|0\rangle=\int^1_0dxe^{ixp\cdot z}(\Phi_K(x))_{\alpha\beta},
\end{eqnarray}
where the wave function of the $K$ meson is given by
\begin{eqnarray}
    &&\Phi_K(x)=\frac{i}{\sqrt{2N_c}}\notag\\
    &&\times \gamma_5\big[\rlap{/}{p_2}\phi_K^A(x) +m_0^K\phi_K^P(x)-m_0^K(\rlap{/}{v}\rlap{/}{n}-1)\phi_K^T(x)\big],
\end{eqnarray}
where $m_0^K=1.6$ GeV is the chiral mass of the $K$ meson. The kaon meson DAs up to twist-three are
\begin{align}
    \phi_K^A(x)=&\frac{f_K}{2\sqrt{{2N_c}}}6x(1-x)\Bigg[1+a_1^KC_1^{3/2}(t) \notag\\
    &+a_2^KC_2^{3/2}(t)+a_4^KC_4^{3/2}(t)\Bigg], \notag\\
    \phi_K^P(x)=&\frac{f_K}{2\sqrt{{2N_c}}}\Bigg[1+\left(30\eta_3-\frac{5}{2}\rho_K^2\right)
    C_2^{1/2}(t) \notag\\
    &-3\left(\eta_3\omega_3+\frac{9}{20}\rho_K^2\left(1+6a_2^K\right)\right)
    C_4^{1/2}(t)\Bigg], \notag\\
    \phi_K^T(x)=&\frac{f_K}{2\sqrt{{2N_c}}}(1-2x)\Bigg[1+6\Big(5\eta_3
    -\frac{1}{2}\eta_3\omega_3 \notag\\
    &-\frac{7}{20}\rho_K^2-\frac{3}{5}\rho_K^2a_2^K\Big)\left(1-10x+10x^2\right)\Bigg],
\end{align}
with $\rho_K=M_K/m_0^K$, where $M_K$ and $f_K$ are the mass and decay constant of the $K$ meson, respectively. The Gegenbauer polynomials are~\cite{Ball:2004ye, Ball:2006wn}
\begin{align}
    C_1^{3/2}(t)&=3t,\quad C_2^{1/2}(t)=\frac{1}{2}(3t^2-1), \notag\\
    C_2^{3/2}(x)&=\frac{3}{2}(5t^2-1), \notag\\
    C_4^{1/2}(t)&=\frac{1}{8}(3-30t^2+35t^4), \notag\\ 
    C_4^{3/2}(t)&=\frac{15}{8}(1-14t^2+21t^4),
\end{align}
with $t=2x-1$, and the Gegenbauer coefficients are~\cite{Ball:2004ye, Ball:2006wn}
\begin{align}
    a_1^K&=0.06, \quad a_2^{K}=0.25\pm 0.15,\quad a_4^{K}=-0.015,\notag\\
    \eta_3^{K}&=0.015, \quad \omega_3^{K}=-3.0.
\end{align}
As in most current works within the PQCD approach, we  incorporate the intrinsic $b$ dependence solely in the LCDA of the $B$ meson,  while neglecting it in the LCDAs of the $K$ meson. 
This is because, in the $B$ meson, the light quark predominantly carries  the longitudinal momentum fraction,  whereas the bottom quark remains nearly at rest.
Consequently, transverse momentum effects become significant, necessitating their explicit inclusion in the LCDA.
In contrast, for the $K$ meson,  the two light quarks  exhibit comparable momentum fractions, and the transverse momentum effects are negligible, justifying their omission in the LCDA.
Previous studies have substantiated this point, as elaborated in Refs.~\cite{Li:1994cka,Li:1995jr,Li:1994iu}. 
However, it should be noted that although the intrinsic $b$ dependence in the LCDAs of the $K$ meson can be neglected, the transverse momentum dependence in the hard scattering kernel must be considered.
Meanwhile, we note that in the calculation of electromagnetic form factors of the $K$ meson, the intrinsic $b$ dependence of the $K$ meson LCDA should be taken into account. For a detailed discussion, see Refs.~\cite{Chai:2024tss,Chai:2025xuz} and the references therein \cite{Jakob:1993iw,Li:2009pr,Kroll:2010bf}.
In addition to this definition of the wave function, we have noticed that another definition of the transverse-momentum-dependent wave function has recently been proposed~\cite{Li:2014xda}.

\section{theoretical Calculations}\label{sec:smcalc}
\begin{figure}[htbp]
    \centering
    \includegraphics[width=1\linewidth]{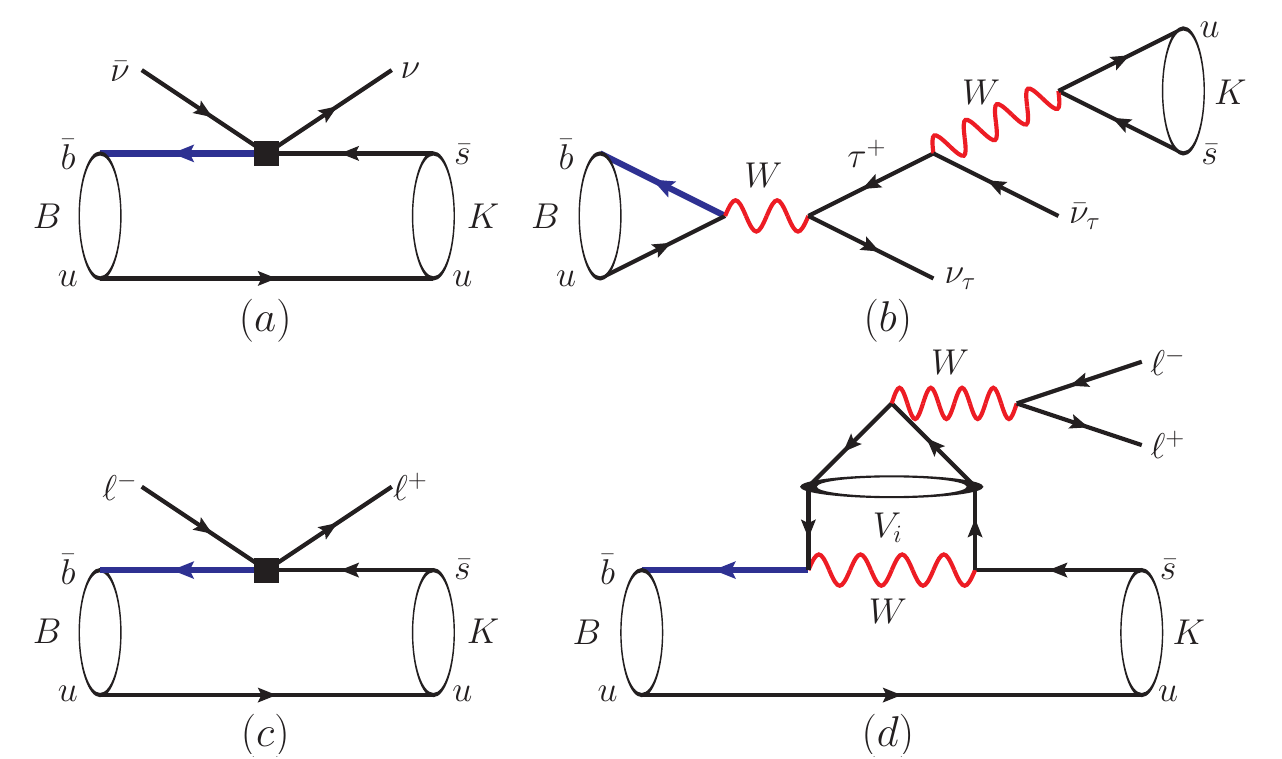}
    \caption{\label{fig:sdld} The Feynman diagrams for $B^+\to K^+$ FCNC processes. (a), (b) Diagram of $B^+\to K^+ \nu\bar \nu$ (c), (d) $B^+\to K^+ \ell^+ \ell^-$ including (a), (c) SD and (b), (d) LD contribution. }
\end{figure}

Utilizing the provided LCDAs, we can analytically compute the semileptonic FCNC decay process of  $B^+ \to K^+$. The considered Feynman diagrams are illustrated in Fig.~\ref{fig:sdld}. This figure depicts both the short-distance (SD) interactions, as represented by the low-energy effective Hamiltonian, and the long-distance (LD) interactions.

The low-energy effective Hamiltonian of $b \to s \nu\bar{\nu}$ transitions can be written as~\cite{Buchalla:1995vs}
\begin{align}\label{eq:heff}
    \mathcal{H}_{\mathrm{eff}}=&
    -\frac{4G_F}{\sqrt{2}}\frac{\alpha_{em}}{2\pi} V_{tb}V_{ts}^* C^{LL}_{\mathrm{SM}}\mathcal{O}^{LL}_V,
\end{align}
where $G_F$ is the Fermi coupling constant, $\alpha_{em}$ is the coupling constant of electromagnetic interaction, $V_{ij}$ are the Cabibbo-Kobayashi-Maskawa (CKM) matrix elements, and $C$ is the Wilson coefficient. The four-fermion operator is
\begin{align}
    \mathcal{O}^{LL}_{V}&=(\bar{s}\gamma^{\mu}P_{L}b)(\bar{\nu}\gamma_{\mu}P_{L}\nu), 
\end{align}
with
\begin{align}
    C^{LL}_{\mathrm{SM}}=-X_t/\sin^2(\theta_W)=-6.32(7).
\end{align}
Here $X_t=1.462(17)$. It includes the NLO QCD corrections~\cite{Buchalla:1993bv,Buchalla:1998ba,Misiak:1999yg} and the two-loop electroweak contributions~\cite{Brod:2010hi}.

For $b\to s\ell^{+}\ell^{-}$ processes, the low-energy effective Hamiltonian is  
\begin{align}
    \mathcal{H}_{\mathrm{eff}}=&-\frac{4G_F}{\sqrt{2}}V_{tb}V_{ts}^*
    \Bigg\{\Big[C_1(\mu)\mathcal{O}_1^c(\mu)+C_2(\mu)\mathcal{O}_2^c(\mu) \notag\\
    &+\sum_{i=3}^{10}C_i(\mu)\mathcal{O}_i(\mu)\Big]
    +\lambda_u\Big[C_1(\mu)(\mathcal{O}_1^c(\mu) \notag\\
    &-\mathcal{O}_1^u(\mu))+C_2(\mu)(\mathcal{O}_2^c(\mu)-\mathcal{O}_2^u(\mu))\Big]\Bigg\},
\end{align}
where $\lambda_u=V_{ub}V_{us}^*/(V_{tb}V_{ts}^*)$. $C_i(\mu)$ are Wilson coefficients, $\mathcal{O}_i(\mu)$ are four fermion operators, and they all depend on the renormalization scale $\mu$. For notational simplicity, we will omit the symbol $\mu$ in the subsequent discussion.
\begin{align}
    \mathcal{O}_{1}^{c}&=(\bar{s}_{\alpha}\gamma^{\mu}P_{L}c_{\beta})(\bar{c}_{\beta}\gamma_{\mu}P_{L}b_{\alpha}), \notag\\
    \mathcal{O}_{2}^{c}&=(\bar{s}_{\alpha}\gamma^{\mu}P_{L}c_{\alpha})(\bar{c}_{\beta}\gamma_{\mu}P_{L}b_{\beta}), \notag\\
    \mathcal{O}_{1}^{u}&=(\bar{s}_{\alpha}\gamma^{\mu}P_{L}u_{\beta})(\bar{u}_{\beta}\gamma_{\mu}P_{L}b_{\alpha}), \notag\\
    \mathcal{O}_{2}^{u}&=(\bar{s}_{\alpha}\gamma^{\mu}P_{L}u_{\alpha})(\bar{u}_{\beta}\gamma_{\mu}P_{L}b_{\beta}), \notag\\
    \mathcal{O}_{3}&=(\bar{s}_{\alpha}\gamma^{\mu}P_{L}b_{\alpha})\sum_{q}(\bar{q}_{\beta}\gamma_{\mu}P_{L}q_{\beta}), \notag\\
    \mathcal{O}_{4}&=(\bar{s}_{\alpha}\gamma^{\mu}P_{L}b_{\beta})\sum_{q}(\bar{q}_{\beta}\gamma_{\mu}P_{L}q_{\alpha}), \notag\\
    \mathcal{O}_{5}&=(\bar{s}_{\alpha}\gamma^{\mu}P_{L}b_{\alpha})\sum_{q}(\bar{q}_{\beta}\gamma_{\mu}P_{R}q_{\beta}), \notag\\
    \mathcal{O}_{6}&=(\bar{s}_{\alpha}\gamma^{\mu}P_{L}b_{\beta})\sum_{q}(\bar{q}_{\beta}\gamma_{\mu}P_{R}q_{\alpha}), \notag\\
    \mathcal{O}_{7}&=\frac{em_b}{16\pi^2}\bar{s}\sigma^{\mu\nu}P_RbF_{\mu\nu}, \notag\\
    \mathcal{O}_{8}&=\frac{gm_b}{16\pi^2}\bar{s}\sigma^{\mu\nu}T^{a}P_RbG_{\mu\nu}^{a}, \notag\\
    \mathcal{O}_{9}&=\frac{\alpha_{em}}{4\pi}(\bar{s}\gamma^{\mu}P_Lb)(\bar{\ell}\gamma_{\mu}\ell), \notag\\
    \mathcal{O}_{10}&=\frac{\alpha_{em}}{4\pi}(\bar{s}\gamma^{\mu}P_Lb)(\bar{\ell}\gamma_{\mu}\gamma_{5}\ell),
\end{align}
$F_{\mu\nu}$ and $G_{\mu\nu}^a$ are the electromagnetic and chromomagnetic tensors, respectively. $T^a$ are the generators of the $SU(3)_C$ group.



Based on the effective Hamiltonian, the transition amplitude for the $B^+ \to K^+$ FCNC process comprises both hadronic and leptonic matrix elements. The hadronic matrix element can be parametrized by the form factors $F_+(q^2)$, $F_0(q^2)$, and $F_T(q^2)$ as
\begin{align}
   & \langle K(p_2)|\bar{s}(0)\gamma_{\mu}b(0)|B(p_1)\rangle
    =\left[\frac{M_B^2-M_K^2}{q^2}q_{\mu}\right]F_0(q^2) \notag\\
   &\qquad\qquad \qquad+\left[(p_1+p_2)_{\mu}-\frac{M_B^2-M_K^2}{q^2}q_{\mu}\right]F_+(q^2),\notag\\
    & \left\langle K(p_2)|\bar{s}(0)\sigma_{\mu \nu}b(0)|B(p_1)\right\rangle
   =\frac{2i[p_{2\mu}q_{\nu}-q_{\mu}p_{2\nu}]}{M_B+M_K}F_T(q^2).
\end{align}
It is straightforward to observe that setting $q^2=0$, $F_+(0)$ results in $F_0(0)$. 



For $B$ decay processes, LO calculations are insufficiently precise for testing the SM. NLO corrections have been extensively considered in semileptonic $B$ decays.
The NLO $B \to \pi$ transition form factors at twist-two are provided in Ref.~\cite{Li:2012nk}. Utilizing these results, the form factors for $B \to K$, $B_s \to K$, and $B \to \pi$ transitions were calculated in Ref.~\cite{Wang:2012ab}. These NLO contributions amount to approximately 20\%. In addition to higher-order corrections, higher-twist effects have also been evaluated in recent studies.  The NLO twist-three contributions to $B \to \pi$ transition form factors have been considered in Ref.~\cite{Cheng:2014fwa}. It was found that the NLO twist-three and twist-two contributions are of similar magnitude but opposite sign, resulting in significant cancellation between them~\cite{Cheng:2014fwa}. Consequently, the NLO twist-three contribution must be accounted for in this study. Given that the behavior of  $B \to K$ processes is analogous to that of $B \to \pi$ transitions, by following the methodology outlined in Refs.~\cite{Li:2012nk, Cheng:2014fwa}, we can readily recalculate the $B \to K$ form factors at NLO. The NLO PQCD factorization formulas for $F_0(q^2)$, $F_+(q^2)$, and $F_T(q^2)$ are
\begin{eqnarray}
    F_0(q^2)|_{\mathrm{NLO}}&=&8\pi C_FM_B^2\int dx_1dx_2\int b_1db_1b_2db_2 \notag\\
    &&\times\phi_B(x_1,b_1)\big\{(2-\eta)r_K[\phi_K^P(x_2)-\phi_K^T(x_2)] \notag\\
    &&\times\alpha_s(t_1)h(x_1,x_2,b_1,b_2)e^{-S_{BK}(t_1)}S_t(x_2) \notag\\
    &&+\eta[(1+x_2\eta)(1+F_{\mathrm{T}2}^{(1)}(x_i,\mu,t,q^2))\phi_K^A(x_2) \notag\\
    &&-2r_Kx_2\phi_K^P(x_2)+2r_K\left(1/\eta-x_2\right)\phi_K^T(x_2)] \notag\\ 
    &&\times\alpha_s(t_1)h(x_1,x_2,b_1,b_2)e^{-S_{BK}(t_1)}S_t(x_2) \notag\\
    &&+\eta[2r_K(1+F_{\mathrm{T}3}^{(1)}(x_i,\mu,t,q^2))\phi_K^P(x_2)]\alpha_s(t_2) \notag\\
    &&\times h(x_2,x_1,b_2,b_1)e^{-S_{BK}(t_2)}S_t(x_1)\big\},
\end{eqnarray}
\begin{eqnarray}
    F_+(q^2)|_{\mathrm{NLO}}&=&8\pi C_FM_B^2\int dx_1dx_2\int b_1db_1b_2db_2 \notag\\
    &&\times\phi_B(x_1,b_1)\big\{r_K[\phi_K^P(x_2)-\phi_K^T(x_2)] \notag\\
    &&\times\alpha_s(t_1)h(x_1,x_2,b_1,b_2)e^{-S_{BK}(t_1)}S_t(x_2) \notag\\
    &&+[(1+x_2\eta)(1+F_{\mathrm{T}2}^{(1)}(x_i,\mu,t,q^2))\phi_K^A(x_2) \notag\\
    &&-2r_Kx_2\phi_K^P(x_2)+2r_K\left(1/\eta-x_2\right)\phi_K^T(x_2)] \notag\\ 
    &&\times\alpha_s(t_1)h(x_1,x_2,b_1,b_2)e^{-S_{BK}(t_1)}S_t(x_2) \notag\\
    &&+[2r_K(1+F_{\mathrm{T}3}^{(1)}(x_i,\mu,t,q^2))\phi_K^P(x_2)]\alpha_s(t_2) \notag\\
    &&\times h(x_2,x_1,b_2,b_1)e^{-S_{BK}(t_2)}S_t(x_1)\big\},
\end{eqnarray}
\begin{eqnarray}
    F_T(q^2)|_{\mathrm{NLO}}&=&8\pi C_FM_B^2(1+r)\int dx_1dx_2\int b_1db_1b_2db_2 \notag\\
    &&\times\phi_B(x_1,b_1)\big\{[(1+F_{\mathrm{T}2}^{(1)}(x_i,\mu,t,q^2))\phi_K^A(x_2) \notag\\
    &&-r_Kx_2\phi_K^P(x_2)+r_K\left(2/\eta+x_2\right)\phi_K^T(x_2)] \notag\\ 
    &&\times\alpha_s(t_1)h(x_1,x_2,b_1,b_2)e^{-S_{BK}(t_1)}S_t(x_2) \notag\\
    &&+[2r_K(1+F_{\mathrm{T}3}^{(1)}\left(x_i,\mu,t,q^2\right))\phi_K^P(x_2)]\alpha_s(t_2) \notag\\
    &&\times h(x_2,x_1,b_2,b_1)e^{-S_{BK}(t_2)}S_t(x_1)\big\},
\end{eqnarray}
with color factor $C_F=4/3$, $r=M_K/M_B$, and $r_K=m_0^K/M_B$. In this context, we have omitted the terms proportional to $x_1$, as they are power suppressed. The running coupling constant $\alpha_s$, hard function $h$, Sudakov exponent $S_{BK}$, and threshold resummation factor $S_t$ can be found in the Appendix. NLO hard kernel $F_{\mathrm{T}2}^{(1)}\left(x_i,\mu,t,q^2\right)$ and $F_{\mathrm{T}3}^{(1)}\left(x_i,\mu,t,q^2\right)$ are also listed in the Appendix.

\section{Numerical results}\label{sec:results}
After constructing the factorization formula for the form factors, we can compute the required form factors using the provided input parameters. The input parameters used in our calculations are summarized as follows. The masses are taken as (in units of GeV)~\cite{ParticleDataGroup:2024cfk}
\begin{align}
    M_B&=5.280, \; M_K=0.494,\; m_{\tau}=1.777, \; m_W=80.37, \notag\\
    m_t&=172.57,\; m_b=4.209, \; m_c=1.272, 
\end{align}
where we omit the mass of $u$, $d$, and $s$ quarks, as well as the $e$ and $\mu$ leptons since they are too light.

The lifetime (in units of picoseconds) and decay constant (in units of GeV) are~\cite{ParticleDataGroup:2024cfk}
\begin{eqnarray}
    \tau_{B}=1.638, \; {\tau_{\tau}=0.29}, \; f_B=0.21, \; f_K=0.16.
\end{eqnarray}
For the CKM matrix elements, we adopt~\cite{ParticleDataGroup:2024cfk}
\begin{align}
    |V_{ub}|&=(3.82\pm0.20)\times10^{-3}, \;|V_{us}|=0.22431\pm0.00085,\notag\\
    |V_{tb}|&=1.010\pm0.027, \; |V_{ts}|=(41.5\pm0.9)\times10^{-3}. 
\end{align}
Wilson Coefficients at $\mu=m_b$~\cite{Blake:2016olu},
\begin{align}
    \begin{array}{lll}
    C_1=-0.294,    & C_2=1.017,   & C_3=-0.0059, \\
    C_4=-0.087,    & C_5=0.0004,  & C_6=0.0011,  \\
    C_7=-0.2957,   & C_8=-0.1630, & C_9=4.114,   \\   
    C_{10}=-4.193. &              &
    \end{array}
\end{align}
Other parameters~\cite{ParticleDataGroup:2024cfk} are
\begin{align}
    &G_F=1.1663788\times10^{-5} \mathrm{GeV}^{-2}, \; \sin^2(\theta_W)=0.231,\notag\\
    &1/\alpha_{em}(M_Z)=127.952,\; \gamma=0.5772156649.
\end{align}

\subsection{Form factors}
Using the input parameters and factorization formula presented above, we can estimate the numerical results for the form factors, which are summarized in Tables~\ref{tab:form0} and~\ref{tab:data}. In Table~\ref{tab:form0}, we detail the contributions from each order and twist. It can be observed that the NLO twist-two and twist-three contributions are approximately equal in magnitude but have opposite signs, leading to an overall enhancement of only about  $3\% \sim 4\%$ relative to the full LO contribution. This finding aligns with the conclusions drawn in Ref.~\cite{Cheng:2014fwa}. In Table~\ref{tab:data}, we compare our results with those from other theoretical studies. The results show consistency among each other within the uncertainties. Within the PQCD framework, the primary sources of uncertainty stem from the shape parameter $\omega_B$ in the $B$ meson DA and the Gegenbauer coefficients of the final-state meson wave functions. Taking into account the uncertainties associated with the input parameters, the total error in our calculations is approximately 20\%.

\begin{table}[htbp]
\centering
\caption{\label{tab:form0}Central values of $B^+ \to K^+$ transition form factors at $q^2=0$, the label LO, LO+NLO(T-2), LO+NLO(T-3), and LO+NLO mean the full LO contribution, the full LO plus NLO twist-two contribution, the full LO plus NLO twist-three contribution, and the total contribution of LO and NLO, respectively.}
\begin{ruledtabular}
\begin{tabular}{lcccc} 
$ $         & LO      & LO+NLO(T-2)    & LO+NLO(T-3)    & LO+NLO         \\ 
\hline
$F_0(0)$    & $0.326$ & $0.381(+17\%)$ & $0.284(-13\%)$ & $0.340(+4\%)$  \\  
$F_+(0)$    & $0.326$ & $0.381(+17\%)$ & $0.284(-13\%)$ & $0.340(+4\%)$  \\
$F_T(0)$    & $0.345$ & $0.399(+16\%)$ & $0.299(-13\%)$ & $0.354(+3\%)$  \\ 
\end{tabular} 
\end{ruledtabular}
\end{table}

\begin{table*}[htbp]
\centering
\caption{\label{tab:data}Results of $B^+ \to K^+$ transition form factors at $q^2=0$ obtained by using different theories. Results of $B^+ \to K^+\ell\bar{\ell}$ branching ratios obtained by using different theories. The data in the fourth, sixth, and seventh column include both SD and LD contributions, while the other data include only SD contributions. The errors in our results originate from form factors. We added the errors mentioned in the references in quadrature. SD results of $B^+ \to K^+\nu\bar{\nu}$ branching ratios $(10^{-6})$ for various $q^2$ bins obtained by using different theories.}
\begin{ruledtabular}
\begin{tabular}{lcccccc} 
$ $      & PQCD(previous)\cite{Wang:2012ab}    & QCDSR\cite{Ball:2004ye,Buras:2014fpa,Tian:2024ubt} & LQCD\cite{Parrott:2022rgu,Parrott:2022zte} & SCET\cite{Cui:2022zwm} & This work       & Data     \\ 
\hline
$F_+(0)$ & $0.310\pm0.054$                     & $0.331\pm0.041$                                    & $0.332\pm0.012$                            & $0.325\pm0.085$        & $0.340\pm0.059$ & $\cdots$ \\
$F_0(0)$ & $0.310\pm0.054$                     & $0.331\pm0.041$                                    & $0.332\pm0.012$                            & $0.325\pm0.085$        & $0.340\pm0.059$ & $\cdots$ \\  
$F_T(0)$ & $0.340\pm0.062$                     & $0.358\pm0.037$                                    & $0.332\pm0.024$                            & $0.351\pm0.097$        & $0.354\pm0.066$ & $\cdots$ \\ 
\hline 
$\mathcal{B}(B^+\to K^+\nu\bar{\nu})(10^{-6})$ &$4.42^{+1.66}_{-1.36}$ & $4.135_{-0.655}^{+0.820}$  & $5.67\pm0.38$  & $5.239^{+0.311}_{-0.281}$ & {$6.69\pm0.66$} & $23\pm7$    \\ 
$\mathcal{B}(B^+\to K^+\ell^+\ell^-)(10^{-7})$ &$5.50^{+2.06}_{-1.69}$ & $6.633_{-1.070}^{+1.341}$  & $7.04\pm0.55$  & $\cdots$                  & {$8.87\pm0.99$} & $4.7\pm0.5$ \\  
\hline
$q^2\in(0,4)$          & $\cdots$ & $0.93\pm0.15$                  & $1.189\pm0.097$ & $1.282^{+0.087}_{-0.080}$                    & {$1.32\pm0.15$} & $\cdots$ \\    
$q^2\in(4,8)$          & $\cdots$ & $0.92\pm0.12$                  & $1.155\pm0.090$ & $1.224^{+0.076}_{-0.069}$                    & {$1.36\pm0.16$} & $\cdots$ \\    
$q^2\in(8,12)$         & $\cdots$ & $0.86\pm0.10$                  & $1.071\pm0.084$ & $1.112^{+0.066}_{-0.060}$                    & {$1.29\pm0.18$} & $\cdots$ \\ 
$q^2\in(12,16)$        & $\cdots$ & $0.71\pm0.08$                  & $0.905\pm0.072$ & $0.916^{+0.053}_{-0.048}$                    & {$1.08\pm0.15$} & $\cdots$ \\    
$q^2\in(16,20)$        & $\cdots$ & \multirow{2}{*}{$0.55\pm0.06$} & $0.597\pm0.048$ & $\multirow{2}{*}{$0.705^{+0.040}_{-0.036}$}$ & {$0.67\pm0.07$} & $\cdots$ \\   
$q^2\in(20,q_{max}^2)$ & $\cdots$ &                                & $0.127\pm0.011$ &                                              & {$0.13\pm0.06$} & $\cdots$ \\  
\end{tabular}
\end{ruledtabular}
\end{table*}

While PQCD offers reliable predictions, it is particularly effective for calculations in the low-$q^2$ region. Conversely, LQCD provides more precise results in the high-$q^2$ region. Hence, we follow the approach outlined in Ref.~\cite{Jin:2020jtu}, employing PQCD for the low-$q^2$region and integrating LQCD results for the high-$q^2$ region to model the form factors over the full kinematic range. The $q^2$ dependence of the form factors is described using the Bourrely-Caprini-Lellouch (BCL) parametrization~\cite{Bourrely:2008za}
\begin{align}
    z(q^2)=\frac{\sqrt{t_+-q^2}-\sqrt{t_+-t_0}}{\sqrt{t_+-q^2}+\sqrt{t_+-t_0}},
\end{align}
with $t_\pm=(M_B \pm M_K)^2$ and $t_0=t_+(1-\sqrt{1-t_-/t_+})$, then the form factor can be expressed as
\begin{align}
    F_i(q^2)=P_i(q^2)\sum_{k}\alpha_k^i[z(q^2)-z(0)]^k,
\end{align}
$P_i(q^2)=(1-q^2/m_{Ri}^2)^{-1}$ is a simple pole, and $m_{R0}\to\infty$, $m_{R+,RT}=5.415$ GeV~\cite{Bharucha:2015bzk}.

\begin{figure*}[htbp]
    \centering
    \includegraphics[width=1\linewidth]{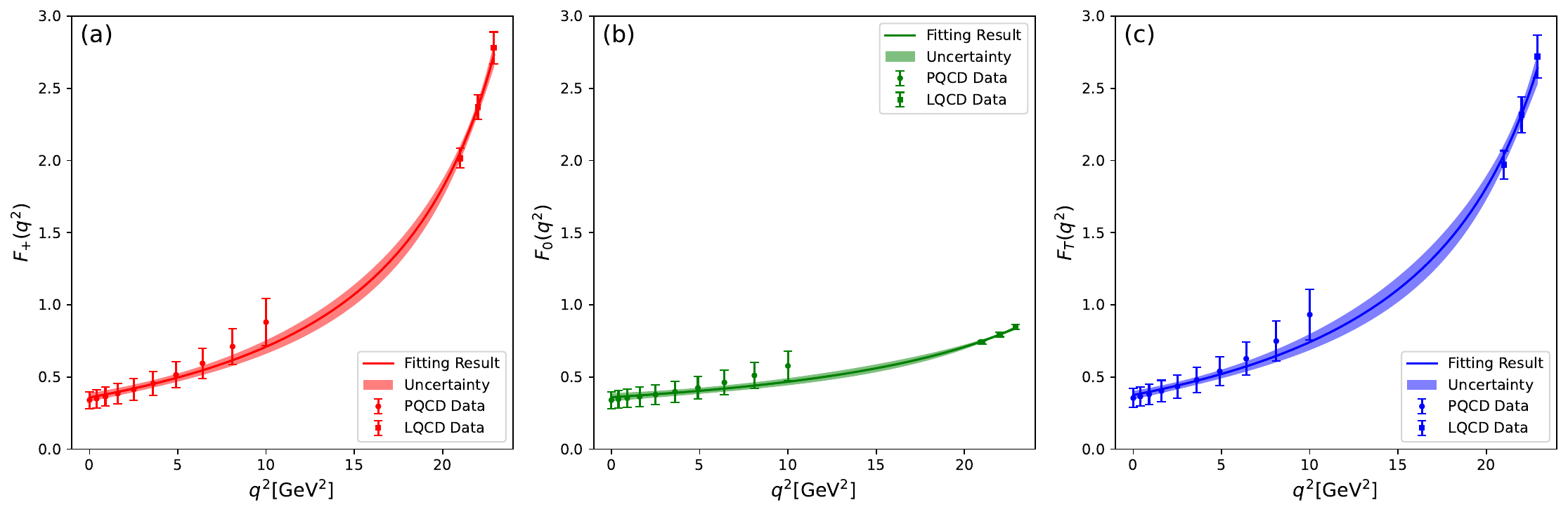}
    \caption{The $q^2$ dependence of the form factors in our fit: (a) $F_+(q^2)$, (b) $F_0(q^2)$, and (c) $F_T(q^2)$.}
    \label{fig:fit}
\end{figure*}

The parameters $\alpha^i_n$, tabulated in  Table~\ref{tab:form2}, are derived from the fitting procedure. Utilizing these fitted parameters, we illustrate the $q^2$ dependence of the form factors in Fig.~\ref{fig:fit}.  
As previously mentioned, the PQCD approach for calculating form factors is predominantly applicable in the large-recoil region, i.e., the low-momentum-transfer region where $q^2 \to 0$. 
We consider that $q^2$ can reach around 10 GeV$^2$, as the previous PQCD studies have adopted similar values~\cite{Wang:2012ab}. We note that as $q^2$ increases, the reliability of the data gradually decreases, which necessitated denser sampling within smaller $q^2$ intervals.

\begin{table}[htbp]
\centering
\caption{\label{tab:form2}{The fitted parameters with BCL parametrization.}}
\begin{ruledtabular}
\begin{tabular}{lccc}   
$ $    & $\alpha_0$  & $\alpha_1$   & $\alpha_2$  \\ 
\hline
$F_+$  & $0.358(28)$ & $-1.42(65)$  & $-2.1(2.2)$ \\  
$F_0$  & $0.360(27)$ & $-1.01(52)$  & $2.3(1.8)$  \\
$F_T$  & $0.375(30)$ & $-1.55(77)$  & $-2.9(2.6)$ \\ 
\end{tabular}  
\end{ruledtabular}
\end{table}

\subsection{$B^+\to K^+\nu\bar{\nu}$}
\begin{figure}[htbp]
    \centering
    \begin{minipage}[t]{0.49\linewidth}
		\centering
		\includegraphics[width=1\columnwidth]{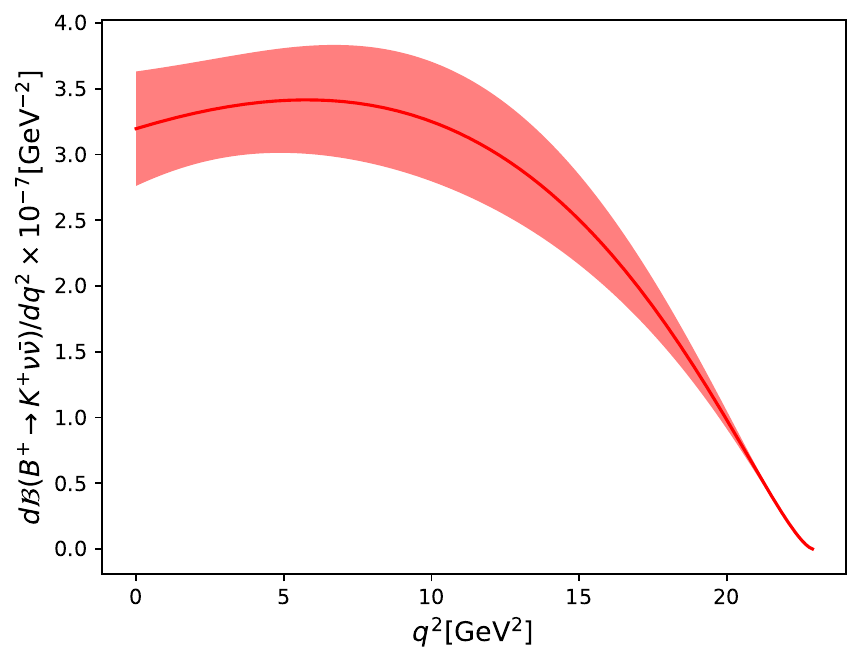}
	\end{minipage}
    \begin{minipage}[t]{0.49\linewidth}
		\centering
		\includegraphics[width=1\columnwidth]{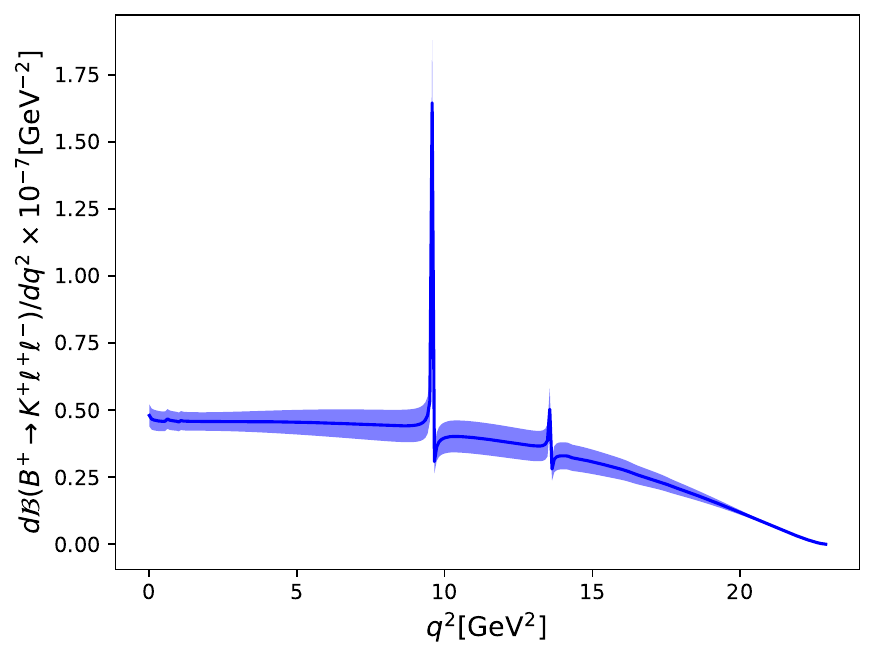}
	\end{minipage}
    \caption{The $q^2$ dependence of the differential branching ratio. }
    \label{fig:q2d}
\end{figure}

Based on the form factors we have fitted,  we can give the decay branching ratios as
\begin{align}\label{eq:dgamma}
    & \mathcal{B}(B^+ \to K^+\nu\bar{\nu})|_{\mathrm{SD}} \notag\\
    & =\tau_{B}\int dq^2\frac{3G_F^2\alpha_{em}^2\lambda_K^{3/2}}
    {768\pi^5M_B^3}|V_{tb}V_{ts}^*|^2|C^{LL}_{\mathrm{SM}}|^2|F_+(q^2)|^2,
\end{align}
where $\lambda_K=\lambda(M_B^2,M_K^2,q^2)$ is the $\mathrm{K\ddot{a}ll\acute{e}n}$ function as
\begin{align}
    \lambda(a,b,c)=a^2+b^2+c^2-2ab-2bc-2ac.
\end{align}
The SD contribution is {$5.85(65)\times10^{-6}$}. For comparing with the experimental data, we also calculate the LD contribution~\cite{Kamenik:2009kc}
\begin{align}
    &\mathcal{B}(B^+\to K^+\nu_\tau\bar{\nu}_\tau)|_{\mathrm{LD}}
    =\tau_{B}\frac{|G_F^2V_{ub}V_{us}^*f_{K}f_{B}|^2}{256\pi^3M_{B}^3} \notag\\
    &\quad\quad\quad\times\tau_\tau2\pi m_\tau(M_{B}^2-m_\tau^2)^2(M_{K}^2-m_\tau^2)^2.
\end{align}
The contribution of LD interaction is $8.41(88)\times10^{-7}$ which could account for 13\% of the total $B^+ \to K^+\nu\bar{\nu}$ rates, but it does not seem to be enough to explain the deviation between theory and experiment. Our numerical results, along with other theoretical predictions, are summarized in Table~\ref{tab:data}. Overall, our results are consistent with those from various theoretical approaches, yet they are notably lower than the experimental data reported by Belle II. In Table~\ref{tab:data}, we present the values for different $q^2$ bins for a more detailed comparison. Additionally, we illustrate the $q^2$ dependence of the differential branching ratio in Fig.~\ref{fig:q2d}. 

To clearly see the difference between experimental measurements and our results, we obtain
\begin{align}\label{eq:rknu}
    R_K^\nu\equiv\frac{\mathcal{B}(B^+ \to K^+\bar{\nu}\nu)|_{\mathrm{Exp}}}
    {\mathcal{B}(B^+ \to K^+\bar{\nu}\nu)|_{\mathrm{SM}}}=3.4\pm1.1.
\end{align}
This indicates that the experimental data are 3~-~4 times higher than the theoretical predictions. To account for this significant discrepancy, one possibility is to introduce NP effects. Further analysis is provided in the following section.

\subsection{$B^+\to K^+\ell^{+}\ell^{-}$}
As for the $B^+\to K^+\ell^{+}\ell^{-}$ process, the contributions from $O_{1-6,8}$ which including SD and LD interaction can be added to the effective Wilson coefficients $O_{7,9}^\mathrm{eff}$, then the effective Hamiltonian can be simplified to
\begin{align}
    \mathcal{H}_{\mathrm{eff}}=&-\frac{4G_{F}}{\sqrt{2}}V_{tb}V_{ts}^{*}
    \frac{\alpha_{em}}{4\pi}
    \Bigg\{C_{10}[\bar{s}\gamma^\mu P_Lb][\bar{\ell}\gamma_\mu\gamma_5\ell] \notag\\
    &+C_{9}^{\mathrm{eff}}[\bar{s}\gamma^\mu P_Lb][\bar{\ell}\gamma_\mu\ell] \notag\\
    &-2m_bC_{7}^{\mathrm{eff}}\left[\bar{s}i\sigma^{\mu\nu}
    \frac{q_{\nu}}{q^{2}}P_{R}b\right][\bar{\ell}\gamma_{\mu}\ell]\Bigg\}.
\end{align}
$C_{7}^{\mathrm{eff}}$ and $C_{9}^{\mathrm{eff}}$ are defined as
\begin{align}
    C_7^{\mathrm{eff}}&=C_{7}+C_{b\to s\gamma}^{\prime}, \notag\\
    C_9^{\mathrm{eff}}&=C_9+Y_{\mathrm{pert}}(\hat{s})+Y_{\mathrm{res}}(\hat{s}),
\end{align}
with $\hat{s}=q^2/M_B^2$. The term which contains the contributions of $b\to s\gamma$ is given by
\begin{align}
    C_{b\to s\gamma}^{\prime}=&i\alpha_{s}(m_b)
    \Bigg\{\frac{2}{9}\eta^{14/23}\Big[\frac{x_t(x_t^2-5x_t-2)}{8(x_t-1)^3}\notag\\
    &+\frac{3x_t^2\ln x_t}{4(x_t-1)^4}-0.1687\Big]-0.03C_{2}\bigg\},
\end{align}
where $\eta=\alpha_{s}(m_{W})/\alpha_{s}(\mu)$ and $x_{t}=m_{t}^{2}/m_{W}^{2}$.
The term $Y_{\mathrm{pert}}(\hat{s})$ contains the SD perturbative contributions
\begin{align}
    Y_{\mathrm{pert}}(\hat{s})=&0.124\omega(\hat{s})+g(\hat{m}_c,\hat{s})C_0 \notag\\
    &+\lambda_u\left[g(\hat{m}_c,\hat{s})-g(0,\hat{s})\right](3C_1+C_2) \notag\\
    &-\frac{1}{2}g(0,\hat{s})(C_3+3C_4) \notag\\
    &-\frac{1}{2}g(\hat{m}_b,\hat{s})(4C_3+4C_4+3C_5+C_6) \notag\\
    &+\frac{2}{9}(3C_3+C_4+3C_5+C_6),
\end{align}
with $C_{0}=3C_{1}+C_{2}+3C_{3}+C_{4}+3C_{5}+C_{6}$ and $\hat{m}_{q}=m_{q}/m_{b}$.
$\omega(\hat{s})$ can be defined as
\begin{align}
    \omega(\hat{s})=&-\frac{2}{9}\pi^2+\frac{4}{3}\int_0^{\hat{s}}\frac{\ln(1-u)}{u}du
    -\frac{2}{3}\ln(\hat{s})\ln{(1-\hat{s})} \notag\\
    &-\frac{5+4\hat{s}}{3(1+2\hat{s})}\ln{(1-\hat{s})}
    -\frac{2\hat{s}(1+\hat{s})(1-2\hat{s})}{3(1-\hat{s})^2(1+2\hat{s})}\ln(\hat{s}) \notag\\
    &+\frac{5+9\hat{s}-6\hat{s}^2}{6(1-\hat{s})(1+2\hat{s})}.
\end{align}
The functions $g(\hat{m}_q,\hat{s})$ and $g(0,\hat{s})$ are of the form
\begin{align}
    g(\hat{m}_q,\hat{s})=&-\frac{8}{9}\ln(\hat{m}_q)+\frac{8}{27}+\frac{4}{9}x
    -\frac{2}{9}(2+x)\sqrt{|1-x|} \notag\\
    &\times
    \begin{cases}\ln|\frac{1+\sqrt{1-x}}{1-\sqrt{1-x}}|-i\pi,
    &x<1, \notag\\
    2\arctan\frac{1}{\sqrt{x-1}},
    &x>1,
    \end{cases}\notag\\
    g(0,\hat{s})=&\frac{8}{27}-\frac{8}{9}\ln\frac{m_b}{\mu}-\frac{4}{9}\ln\hat{s}+\frac{4}{9}i\pi,
\end{align}
with $x=4\hat{m}_q^2/\hat{s}$.
The term $Y_{\mathrm{res}}(\hat{s})$ refers to the LD contributions 
\begin{align}
    Y_{\mathrm{res}}(\hat{s})=&-\frac{3\pi}{\alpha_{em}^2}
    \Bigg[C_0\sum_{V_i}\frac{M_{V_i}\mathcal{B}(V_i\to\ell^+\ell^-)\Gamma_{V_i}}
    {q^2-M_{V_i}^2+iM_{V_i}\Gamma_{V_i}} \notag\\
    &-\lambda_ug(0,\hat{s})(3C_1+C_2) \notag\\
    &\times\sum_{V_j}\frac{M_{V_j}\mathcal{B}(V_j\to\ell^+\ell^-)\Gamma_{V_j}}
    {q^2-M_{V_j}^2+iM_{V_j}\Gamma_{V_j}}\Bigg].\label{eq39}
\end{align}
Light vector mesons and charmonium states that may contribute are listed in Table \ref{tab:vector}.
\begin{table}[htbp]
\centering
\caption{\label{tab:vector}Vector mesons masses, widths, and branching ratios to $\ell^+\ell^-$~\cite{ParticleDataGroup:2024cfk}. The first three rows correspond to $V_i$, and the last five rows correspond to $V_j$ in Eq.(\ref{eq39}).}
\begin{ruledtabular}
\begin{tabular}{lccc}  
$V$        & $M_{V}$ (GeV) & $\Gamma_{V}$ (MeV) & $\mathcal{B}(V\to \ell^+\ell^-)$ \\ 
\hline 
$\rho$       & 0.775 & 149   & $4.635\times10^{-5}$ \\
$\omega$     & 0.783 & 8.68  & $7.380\times10^{-5}$ \\
$\phi$       & 1.019 & 4.249 & $2.915\times10^{-4}$ \\
$J/\psi$     & 3.097 & 0.093 & $5.966\times10^{-2}$ \\
$\psi(2S)$   & 3.686 & 0.294 & $7.965\times10^{-3}$ \\
$\psi(3770)$ & 3.774 & 27.2  & $9.6\times10^{-6}$   \\
$\psi(4040)$ & 4.039 & 80    & $1.07\times10^{-5}$  \\
$\psi(4160)$ & 4.191 & 70    & $6.9\times10^{-6}$   \\
\end{tabular}
\end{ruledtabular}
\end{table}

Then the branching ratios can be given by
\begin{align}
    &\mathcal{B}(B^+\to K^+ \ell^+\ell^-) \notag\\
    &=\tau_{B}\int dq^2 \frac{G_{F}^{2}\alpha_{em}^{2}|V_{tb}|^{2}|V_{ts}^{*}|^{2}
    \sqrt{\lambda_K}}{512M_{B}^{3}\pi^{5}}\sqrt{\frac{q^{2}-4m_{l}^{2}}{q^{2}}}
    \frac{1}{3q^{2}} \notag\\
    &\times\bigg[6m_{l}^{2}|C_{10}|^{2}(M_{B}^{2}-M_{K}^{2})^{2}F_{0}^{2}(q^{2})
    +(q^{2}+2m_{l}^{2})\lambda(q^{2}) \notag\\
    &\times\left|C_9^\mathrm{eff}F_+(q^2)
    +\frac{2C_7^\mathrm{eff}(m_b-m_s)F_T(q^2)}{M_B+M_K}\right|^2 \notag\\
    &+|C_{10}|^2(q^2-4m_l^2)\lambda_KF_+^2(q^2)\bigg].
\end{align}

Our results are summarized in Table~\ref{tab:data}. We have updated the prior PQCD calculations, offering more reliable and precise predictions. The $q^2$ dependence branching ratios are also given in Fig.~\ref{fig:q2d}. The two peaks of the branching ratio curve correspond to the possible resonance contribution, such as $J/\Psi$ and $\Psi(2S)$ in Fig.~\ref{fig:sdld}(d). When comparing our results with those from other theoretical approaches and experimental measurements, it is evident that the theoretical predictions tend to be higher than the experimental data. This discrepancy might indicate that certain process contributions are being offset by NP effects.

Analogous to the definition of $R_K^\nu$, we can define 
\begin{align}\label{eq:rkl}
    R_K^\ell\equiv\frac{\mathcal{B}(B^+ \to K^+ \ell^+\ell^-)|_{\mathrm{Exp}}}
    {\mathcal{B}(B^+ \to K^+ \ell^+\ell^-)|_{\mathrm{SM}}}=0.530\pm0.082.
\end{align}
While this discrepancy is less pronounced than that observed for $R_K^\nu$, it nonetheless calls for a theoretical explanation.

\section{Leptoquark effects}\label{sec:npm}
At the bottom-quark mass scale, considering the lepton flavor universality (LFU), the dimension-six NP effective Hamiltonian of $b \to s $ semileptonic FCNC transitions can be written as~\cite{Dorsner:2016wpm,Browder:2021hbl}
\begin{align}
    \mathcal{H}_{\mathrm{eff}}=-\frac{4G_F}{\sqrt{2}}\frac{\alpha_{em}}{2\pi} V_{tb}V_{ts}^*
    \sum_{X,A,B}C^{AB}_X\mathcal{O}^{AB}_X,
\end{align}
with $X=V,S,T$ and $A,B=L,R$, the ten four-fermion operators
\begin{align}
    \mathcal{O}^{AB}_{V}&=(\bar{s}\gamma^{\mu}P_{A}b)(\bar{\ell}\gamma_{\mu}P_{B}\ell),\notag\\
    \mathcal{O}^{AB}_{S}&=(\bar{s}P_{A}b)(\bar{\ell}P_{B}\ell),\notag\\
    \mathcal{O}^{AB}_{T}&=\delta_{AB}(\bar{s}\sigma^{\mu\nu}P_{A}b)(\bar{\ell}\sigma_{\mu\nu}P_{B}\ell).
\end{align}
Here, $\ell$ denotes neutrinos or charged leptons.

Considering the possible right-hand-neutrinos are too heavy to be produced in low-energy processes, the right-hand lepton including $\ell_R$ and $\nu_R$ will not be considered in our work.
Based on the previous studies~\cite{Dorsner:2016wpm}, there are five different LQ candidates that can explain the $B^+\to K^+$ anomaly: $S_1=(\bar{3},1,1/3)$, $\tilde{R}_2=(3,2,1/6)$, $S_3=(\overline{3},3,1/3)$, $V_2=(\overline{3},2,5/6)$, and $U_3=(3,3,2/3)$. Here the numbers in the brackets represent the SM gauge groups $SU(3)_C$, $SU(2)_L$, and $U(1)_Y$ quantum numbers, respectively. The former three mean the scalar LQ, and the latter two stand for vector LQs. The corresponding interactions are 
\begin{eqnarray}
    &&S_1:\;        \mathcal{L}_{S_1}\supset+y_{1ij}^{LL}\bar{Q}_{L}^{Ci,a}S_{1}\epsilon^{ab}L_{L}^{j,b}+\mathrm{H.c.},\nonumber\\
    &&\tilde R_2:\; \mathcal{L}_{\tilde R_2}\supset-\tilde{y}_{2ij}^{RL}\bar{d}_{R}^{i}\tilde{R}_{2}^{a}\epsilon^{ab}L_{L}^{j,b}+\mathrm{H.c.},\nonumber\\
    && S_3:\;       \mathcal{L}_{S_3}\supset+y_{3{ij}}^{LL}\bar{Q}_{L}^{C{i,a}}\epsilon^{ab}(\tau^{k}S_{3}^{k})^{bc}L_{L}^{j,c}+\mathrm{H.c.},\nonumber\\
    && V_2:\;       \mathcal{L}_{V_2}\supset+x_{2ij}^{RL}\bar{d}_{R}^{C}\gamma^{\mu}V_{2,\mu}^{a}\epsilon^{ab}L_{L}^{j,b}+\mathrm{H.c.},\nonumber\\
    && U_3:\;       \mathcal{L}_{U_3}\supset+x_{3{ij}}^{LL}\bar{Q}_{L}^{i,a}\gamma^{\mu}(\tau^{k}U_{3,\mu}^{k})^{ab}L_{L}^{j,b}+\mathrm{H.c.},
\end{eqnarray}
where we adopt the notation for LQs in Ref.~\cite{Dorsner:2016wpm}.
After integrating out the heavy LQ field, we can obtain the corresponding Wilson coefficients as shown in Table~\ref{tab:coef}, by adopting the Fierz transformation into the vector current form. 

Then, the general effective Hamiltonian responsible for the decay $B^+\to K^+\nu\bar\nu$ is written as 
\begin{eqnarray}
    \mathcal{H}_{\mathrm{eff}}=-\frac{4G_F}{\sqrt{2}}\frac{\alpha_{em}}{2\pi}V_{tb}&V_{ts}^*
    &\Big[C_{\mathrm{SM}}^{LL}(\bar{s}\gamma^{\mu}P_{L}b)(\bar{\nu}\gamma_{\mu}P_{L}\nu) \notag\\
    &+&C_V^{LL}(\bar{s}\gamma^{\mu}P_{L}b)(\bar{\nu}\gamma_{\mu}P_{L}\nu)\notag\\
    &+&C_V^{RL}(\bar{s}\gamma^{\mu}P_{R}b)(\bar{\nu}\gamma_{\mu}P_{L}\nu)\Big].
\end{eqnarray}
The corresponding branching ratios are obtained by 
\begin{align}\label{eq:rrr}
    \mathcal{B}(B^+\to K^+\nu\bar{\nu})=&\mathcal{B}(B^+\to K^+\nu\bar{\nu})|_{\mathrm{SM}} \notag\\
    &\times\left|1+\frac{C^{LL}_V+C_V^{RL}}{C^{LL}_{\mathrm{SM}}}\right|^2.
\end{align}
Therefore, the deviation between SM prediction and experimental data can be attributed to the vector current coupling LQ operator associated with  the Wilson coefficients $C_{V}^{LL,RL}$.

\begin{table}[htbp]
\centering
\caption{\label{tab:coef}Tree-level Wilson coefficients of LQ models in $b \to s\nu\bar{\nu}$ and $b \to s\ell^+\ell^-$ \cite{Dorsner:2016wpm}, $\lambda_t=V_{tb}V_{ts}^{*}$, $x,y$ are Yukawa coupling matrix elements.}
\begin{ruledtabular}
\begin{tabular}{lcc}  
                        & $b \to s\nu\bar{\nu}$ & $b \to s\ell^+\ell^-$ \\ 
\hline
$C_V^{LL}(S_1$)         & $\frac{v^{2}}{2M_{\mathrm{LQ}}^{2}}\frac{\pi}{\alpha_{em}\lambda_t}y_{1b}^{LL}y_{1s}^{LL*}$ & $\cdots$\\
$C_V^{RL}(\tilde{R}_2)$ & $-\frac{v^{2}}{2M_{\mathrm{LQ}}^{2}}\frac{\pi}{\alpha_{em}\lambda_t}\tilde{y}_{2b}^{RL}\tilde{y}_{2s}^{RL*}$ & $-\frac{v^{2}}{2M_{\mathrm{LQ}}^{2}}\frac{\pi}{\alpha_{em}\lambda_t}\tilde{y}_{2b}^{RL}\tilde{y}_{2s}^{RL*}$\\ 
$C_V^{LL}(S_3)$         & $\frac{v^{2}}{2M_{\mathrm{LQ}}^{2}}\frac{\pi}{\alpha_{em}\lambda_t}y_{3b}^{LL}y_{3s}^{LL*}$ & $\frac{v^{2}}{M_{\mathrm{LQ}}^{2}}\frac{\pi}{\alpha_{em}\lambda_t}y_{3b}^{LL}y_{3s}^{LL*}$\\
$C_V^{RL}(V_2)$         & $\frac{v^{2}}{M_{\mathrm{LQ}}^{2}}\frac{\pi}{\alpha_{em}\lambda_t}x_{2b}^{RL}x_{2s}^{RL*}$ & $\frac{v^{2}}{M_{\mathrm{LQ}}^{2}}\frac{\pi}{\alpha_{em}\lambda_t}x_{2b}^{RL}x_{2s}^{RL*}$\\
$C_V^{LL}(U_3)$         & $-\frac{2v^{2}}{M_{\mathrm{LQ}}^{2}}\frac{\pi}{\alpha_{em}\lambda_t}x_{3b}^{LL}x_{3s}^{LL*}$ & $-\frac{v^{2}}{M_{\mathrm{LQ}}^{2}}\frac{\pi}{\alpha_{em}\lambda_t}x_{3b}^{LL}x_{3s}^{LL*}$\\ 
\end{tabular}  
\end{ruledtabular}
\end{table}

Comparing the experimental results and our new prediction in Eq.~(\ref{eq:rknu}), we can obtain the viable LQ parameter ranges. As shown in Table~\ref{tab:coef}, the above five different types of LQ particles both can  contribute to $B^+\to K^+\nu\bar\nu$.
For simplicity, we consider only one type of LQ contributing in each physical decay process.
Within a 1$\sigma$ error range, the numerical ranges for the corresponding Wilson coefficients are estimated as
\begin{align}\label{eq:cc}
    &C_V^{LL}(S_1,S_3,U_3)=C_V^{RL}(\tilde{R}_2,V_2)\notag\\
    &=[-7.3,-3.5]\cup[16.1,19.9].
\end{align}
In this context, the identical coefficients across all five LQs  arise from their analogous contributions to Eq.~(\ref{eq:rrr}).

For the $B^+ \to K^+ \ell^+\ell^-$ process, with the addition of NP contributions, the effective Hamiltonian can be written as
\begin{align}
    \mathcal{H}_{\mathrm{eff}}=&-\frac{4G_{F}}{\sqrt{2}}V_{tb}V_{ts}^{*}
    \frac{\alpha_{em}}{4\pi} \notag\\
    &\times\Bigg\{(C_{10}-C_V^{LL}-C_V^{RL})[\bar{s}\gamma^\mu P_Lb][\bar{\ell}\gamma_\mu\gamma_5\ell]    \notag\\
    &+(C_{9}^{\mathrm{eff}}+C_V^{LL}+C_V^{RL})[\bar{s}\gamma^\mu P_Lb][\bar{\ell}\gamma_\mu\ell] \notag\\
    &-2m_bC_{7}^{\mathrm{eff}}\left[\bar{s}i\sigma^{\mu\nu}
    \frac{q_{\nu}}{q^{2}}P_{R}b\right][\bar{\ell}\gamma_{\mu}\ell]\Bigg\}.
\end{align}
In this analysis, we disregard the contributions from right-handed charged leptons for simplicity. Instead, we focus solely on the LQs that contribute concurrently to both charged lepton and neutrino final states, consistent with $SU(2)_L$ gauge invariance. The corresponding Wilson coefficients are provided in Table.~\ref{tab:coef}. Notably, $S_1$ does not contribute to the charged lepton process, leaving only the remaining four LQs as contributors.


With the inclusion of NP contributions, the relationship between the decay branching ratio of $B^+\to K^+\ell^+\ell^-$ and its SM counterpart becomes more complex than suggested by Eq.~(\ref{eq:rrr}). To address this, the differential branching ratio must be expanded based on the distinct structures of the Wilson coefficients, followed by separate integration to derive the following form as
\begin{align}
    &\mathcal{B}(B^+\to K^+\ell^+\ell^-)\times 10^{7}\notag\\
    &=0.249(26)|C_{10}-C_{10}^{\mathrm{NP}}|^2
    +0.249(26)|C_{9}+C_{9}^{\mathrm{NP}}|^2 \notag\\
    &-0.259(22)(C_9+C_{9}^{\mathrm{NP}})+1.35(0.18).
\end{align}
Based on the above equations and Eq.~(\ref{eq:rkl}), the corresponding numerical ranges for the coefficients $C_V^{LL,RL}$ can be determined within the 1$\sigma$ uncertainty range as follows:
\begin{align}\label{eq:cbsll}
    &C_V^{LL}(S_3,U_3)=C_V^{RL}(\tilde{R}_2,V_2)\notag\\
    &=[-7.4,-5.6]\cup[-1.8,-0.8].
\end{align}
Similar to the previous scenario, the four LQs contribute to  NP in an identical manner, resulting in the same coefficient values.



\begin{figure}[htbp!]
    \centering
    \includegraphics[width=1\linewidth]{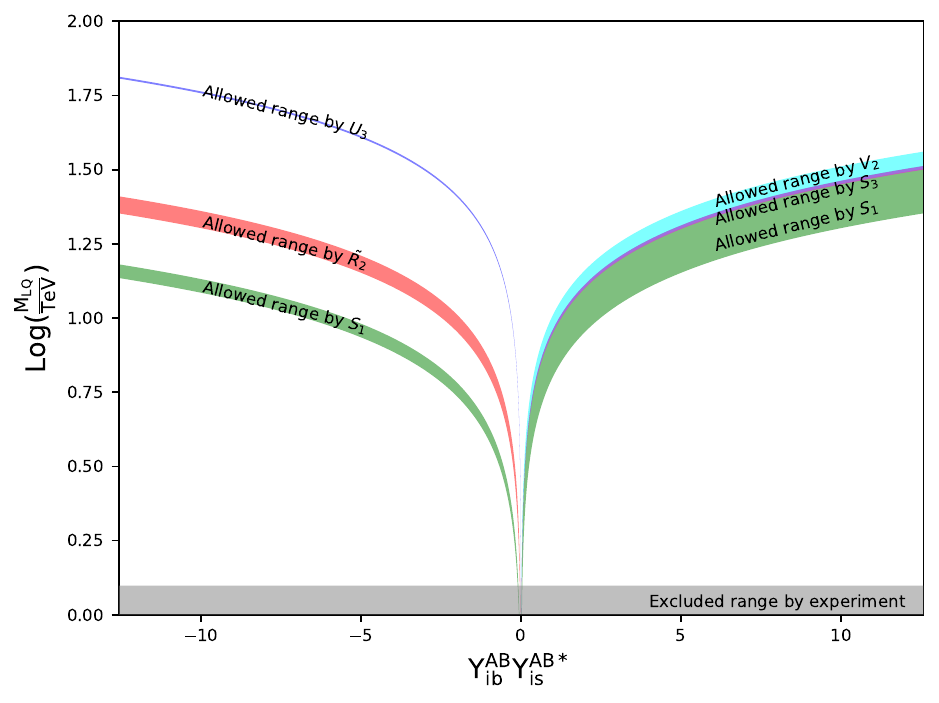}
    \caption{The $B^+\to K^+$ decay allowed region on the LQ mass and the coupling constant ($Y=x,y$) for different LQ scenarios ($S_1, \tilde{R}_2, S_3, V_2, U_3$).
    The different LQ scenarios are shown in different colors. The symbols A and B mean the chirality (L,R).
    Note that the broader viable region for $S_1$ is due to the lack of the $R_K^\ell$ constraints.}
    \label{fig:ym}
\end{figure}

These five LQ models are observed to individually account for the experimental data of the $B^+\to K^+\nu\bar\nu$ and $B^+\to K^+\ell^+\ell^-$ decay channels. By combining the results from both decay channels, the parameter space of the LQs can be further constrained. Although the Wilson coefficients for each type of LQ involved in the above two decays (charged leptons and neutrinos) differ with different factors, as presented in Table.~\ref{tab:coef}, the underlying structure induced by Yukawa couplings and LQ masses remain consistent. Consequently, the parameter space of the LQ mass and the Yukawa coupling matrix elements, which constitute the Wilson coefficient $C^{\mathrm{NP}}$, can be constrained through the FCNC $B^+\to K^+$ decays. The allowed parameter space for the five LQs is presented in Fig.~\ref{fig:ym}. Additionally, the current experimental bounds on the LQ mass from colliders provide the lower limit for the masses of LQs coupled to three generations of leptons, as discussed in Ref.~\cite{ParticleDataGroup:2024cfk}. Considering the requirement of LFU in our analysis, we adopt the average lower bound of 1250 GeV for the LQ mass, as indicated in gray.
For satisfying the perturbative unitarity bound, we choose the Yukawa coupling $Y\leq \sqrt{4\pi}$, which leads to the product  $Y_{ib}^{AB}Y_{is}^{AB*}$ across the entire range $[-4\pi,4\pi]$.


As illustrated in Fig.~\ref{fig:ym}, the semileptonic FCNC decay $B^+\to K^+$ imposes significant constraints on the mass range of LQs, under the assumption that the NP contribution arises exclusively from LQs. For the $S_1$ type LQ, as it does not  contribute to the $B^+\to K^+\ell^+\ell^-$ decay, its allowed parameter space is broader than those of other LQ types. Furthermore, the product $Y^{AB}_{ib}Y^{AB*}_{is}$ spans both positive and negative values. The LQ masses exhibit opposite trends in the positive and negative coupling product ranges, increasing with positive values and decreasing with negative values. For the $\tilde{R}_2$ and $U_3$ type LQs, the product of the Yukawa couplings $Y^{AB}_{ib}Y^{AB*}_{is}$ is restricted to negative values. The allowable masses of these LQs increase as the product $Y^{AB}_{ib}Y^{AB*}_{is}$ decreases.
For the $V_2$ and $S_3$ LQs, only positive values of $Y^{AB}_{ib}Y^{AB*}_{is}$ are viable and the corresponding LQ masses increase as  $Y^{AB}_{ib}Y^{AB*}_{is}$ grows.
Furthermore, requiring the perturbative unitarity bound $|Y^{AB}_{ib}Y^{AB*}_{is}|\leq 4\pi$, we can obtain the LQ allowed upper bound (in TeV)
\begin{align}
    M_{S_1}&\approx 33,\quad
    M_{\tilde{R}_2}\approx 26,\quad
    M_{S_3}\approx 33,\notag\\
    M_{V_2}&\approx 36,\quad
    M_{U_3}\approx 65.
\end{align}

For the above two processes, LQ models can construct the four-fermion interactions, $bs\nu\nu$ or  $bs\ell\ell$.  The interaction structures involve in the flavor changing quark $b\to s$ and  flavor conserving lepton pairs, which can also contribute to other physical processes.
	
The first one is decay channel $B_s\to \ell^+\ell^-$ with the branching ratio as
\begin{align}\label{eq:bsll}
	&R_{B_s\to \ell^+\ell^-}=\frac{\mathcal{B}(B_s\to \ell^+\ell^- )|_\mathrm{Exp}}{\mathcal{B}(B_s\to \ell^+\ell^- )|_\mathrm{SM}}=\left|1-\frac{C_{V}^{LL}-C_{V}^{RL}}{2C_{10}^\mathrm{SM}}\right|^2 .
\end{align}
Here the SM Wilson coefficients are $C_{10}^\mathrm{SM}\approx -4.2$~\cite{Hiller:2014yaa}. Until now, the current strongest bounds on the branching ratio come from $B_s\to \mu^+\mu^-$ with experimental values $3.34\pm0.27$~\cite{CMS:2022mgd} and SM prediction $3.78\pm0.15$~\cite{Buras:2022qip}, which results in $R_{B_s\to \ell^+\ell^-}=0.88\pm 0.08$.

Substituting this result into Eq.~(\ref{eq:bsll}), we can obtain the range of values for the coefficients within 1$\sigma$,
\begin{align}
    &C_V^{LL}(S_3,U_3)=[-16.7,-15.9]\cup[-0.9,-0.1],\notag\\
    &C_V^{RL}(\tilde{R}_2,V_2)=[0.1,0.9]\cup[15.9,16.7].
\end{align}
Comparing  with these intervals in Eq.~(\ref{eq:cc}), we find that there is no overlapping region. This indicates that the single LQ model becomes invalid as a possible explanation of these above two anomalies ($B\to K\nu\bar{\nu},K\ell^+\ell^-$) after considering the bound from the $B_s\to \mu^+\mu^-$ process, which is consistent with the previous studies~\cite{Bause:2023mfe,He:2023bnk}.

The another process is $B_s$-$\bar B_s$ mixing mediated by LQs via the box diagram. 
For the scalar and vector LQs, the corresponding effective Lagrangian is~\cite{Dorsner:2016wpm}
\begin{align}
	& \mathcal{L}^S_\mathrm{eff}=-\frac{1}{128\pi^2M^2_\mathrm{LQ}}\sum_i (y_{ib}y_{is}^*)^2\bar s\gamma^\mu P_R b \bar s\gamma_\mu P_R b  ,\notag\\
	&  \mathcal{L}^V_\mathrm{eff}=\frac{1}{32\pi^2M^2_\mathrm{LQ}}\sum_i (x_{ib}x_{is}^*)^2S_0(x_i,x_i)\bar s\gamma^\mu P_L b \bar s\gamma_\mu P_L b  ,\notag\\
	&  \mathcal{L}_\mathrm{SM}=-\frac{4G_f}{\sqrt{2}}(V_{tb}V_{ts}^*)^2 C^\mathrm{SM}_{bs}\bar s\gamma^\mu P_L b \bar s\gamma_\mu P_L b  .
\end{align}
Here $x_i=m_i^2/M_\mathrm{LQ}^2$,  the SM coefficients are  $C^\mathrm{SM}_{bs}\approx 0.86\times 10^{-3}$~\cite{Becirevic:2024pni} and the Inami–Lim function $S_0(0,0)\sim 0$.
Correspondingly, we obtain the contribution to the mass differences as
\begin{align}\label{eq:mbb}
	& \frac{\Delta m_{B_s}}{\Delta m^\mathrm{SM}_{B_s}}=\left|1+3\frac{(y_{ib}y_{is}^*)^2}{128\pi^2M^2_\mathrm{LQ}}
	\frac{\sqrt{2}}{4G_f (V_{tb}V_{ts}^*)^2C_{bs}^\mathrm{SM}}\right|\;.
\end{align}
Note that we only indicate the contribution from the scalar LQs because the one from vector LQs is approximately around 0 for $m_l^2/M_\mathrm{LQ}^2\sim 0$.
The current experimental and theoretical values are $\Delta m_{B_s}=17.765\pm 0.006$ ps~\cite{ParticleDataGroup:2024cfk} and 
$\Delta m^\mathrm{SM}_{B_s}=18.23\pm 0.63$ ps$^{-1}$~\cite{Albrecht:2024oyn}, which results in the ratio as $0.975\pm0.03$.

\begin{figure}
    \centering
    \includegraphics[width=1\linewidth]{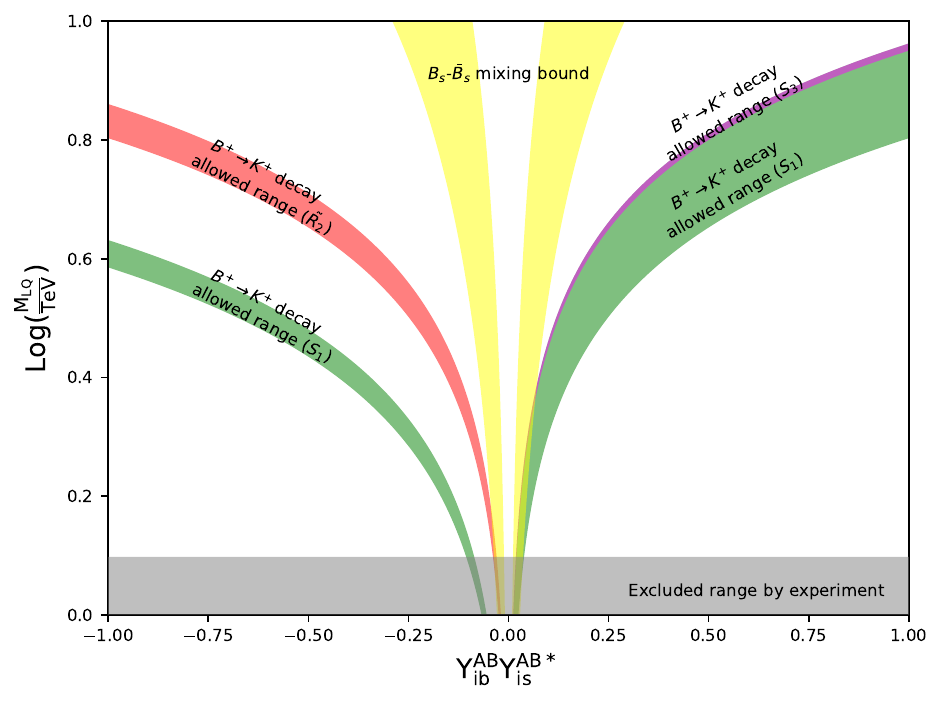}
    \caption{The $B^+\to K^+$ decay allowed region and $B_s$-$\bar B_s$ mixing bound on the scalar LQ models for the scenarios of $S_1$, $\tilde R_2$, and $S_3$. 
    The notations are the same as Fig.~\ref{fig:ym}.
    }
    \label{fig:2sigma}
\end{figure}

Based on this ratio,  we can obtain the  corresponding constraints within $2\sigma$ errors as
\begin{align}
	\frac{{|y_{ib}y_{is}^*|}}{M_\mathrm{LQ}}=
		(1.9\pm1.0)\times 10^{-5} (\mathrm{GeV})^{-2}.
\end{align}
The corresponding bounds within $2\sigma$ are also shown in Fig.~\ref{fig:2sigma}, marked in yellow.
We find that there exists some viable overlapping area for explaining the $B_s$-$\bar B_s$ mixing and $B^+\to K^+$ decay simultaneously, in particular, for $S_1$, $\tilde R_2$, and $S_3$.



\section{Summary}\label{sec:summary}
We have analyzed the $B^+ \to K^+$ transition form factor at the next-to-leading order and twist-three level within the framework of perturbative QCD factorization.
Using the latest lattice QCD data, we performed a comprehensive fit across the full dynamical region, as shown in Fig.~\ref{fig:fit}, which improves the reliability of the form factors.
Based on these fitted form factors, we computed the branching ratio for the $B^+ \to K^+ \nu \bar{\nu}$ and $B^+ \to K^+ \ell^+ \ell^- $ decay.
The $q^2$ dependence of the differential branching ratio is shown in Fig.~\ref{fig:q2d}, allowing for a direct comparison with existing experimental data.
The branching ratios for these two processes are provided as
\begin{eqnarray}
    &&\mathcal{B}(B^+\to K^+\nu\bar\nu)=(6.69\pm0.66)\times 10^{-6},\notag\\
    &&\mathcal{B}(B^+\to K^+\ell^+\ell^-)=(8.87\pm0.99)\times 10^{-7}.
\end{eqnarray}
Our findings show that the Standard Model calculations do not align with the experimental data presented in Table~\ref{tab:data}.
To address this discrepancy between our theoretical predictions and experimental results, we investigated leptoquark models as a  potential  source of new physics.
By incorporating five types of leptoquark models—namely $S_1$, $\tilde{R}_2$, $S_3$, $V_2$, and $U_3$—we were able to reconcile the discrepancy and further constrain the parameter space of these leptoquarks. Constraints on the masses and Yukawa couplings of the leptoquarks are depicted in Fig.~\ref{fig:ym}. 
Our study demonstrates that leptoquarks can significantly influence the semileptonic flavor-changing neutral current $B^+\to K^+$ decays, thereby significantly constraining the viable parameter space for leptoquark masses and Yukawa couplings.
Furthermore,  the decay   $B_s \to \mu^+\mu^-$ and $B_s$-$\bar{B_s}$ mixing impose the stringent bounds, which render   the single leptoquark model untenable as a viable explanation.
We anticipate additional experimental data in the future to reduce uncertainties,
allowing us to rigorously test the validity of leptoquark models and other new physics hypotheses.

\begin{acknowledgments}
We thank Zhou Rui, Ya Li, Su-Ping Jin, Hong-Xin Dong, and Yi-Qi Geng for useful discussion. The work of J. S. is supported by IBS under the project code, IBS-R018-D1. 
The work of R.-L. Z. is supported by NSFC under Grants No. 12322503 and No. 12075124, and by Natural Science Foundation of Jiangsu under Grant No. BK20211267.
The work of Z.-P. X. is supported by NSFC under Grants No.12375088, No. 12335003 and No.12405113. 
\end{acknowledgments}

\appendix
\section{PQCD Functions}\label{appendix}
The NLO running coupling constant~\cite{Li:1994iu}
\begin{align}
    \alpha_s(t)=\frac{\pi}{2\beta_1\hat{t}}
    -\frac{\pi \beta_2}{4\beta_1^3}\frac{\ln2\hat{t}}{\hat{t}^2},
\end{align}
with $\hat{t}=\ln(t/\Lambda)$, $\Lambda$ is QCD the scale, and we adopt $\Lambda_{\mathrm{QCD}}=0.25$.
The hard scales $t$ are chosen as the maximal virtuality of the internal particles,
\begin{align}
    t_1&=\max(M_B\sqrt{\eta x_2},1/b_1,1/b_2), \notag\\ 
    t_2&=\max(M_B\sqrt{\eta x_1},1/b_1,1/b_2).
\end{align}

The hard functions
\begin{align}
    &h(x_1,x_2,b_1,b_2)=K_0(M_B\sqrt{\eta x_1x_2} b_1) \notag\\
    &\times[\theta(b_1-b_2)I_0(M_B\sqrt{\eta x_2} b_2)K_0(M_B\sqrt{\eta x_2} b_1) \notag\\
    &+\theta(b_2-b_1)I_0(M_B\sqrt{\eta x_2} b_1)K_0(M_B\sqrt{\eta x_2} b_2)].
\end{align}

Sudakov exponents $S_{BK}(t)=S_B(t)+S_K(t)$,
\begin{align}
    S_B(t)=&s\left(\frac{M_B}{\sqrt{2}}x_1,b_1\right)+\frac{5}{3}\int_{1/b_1}^t
    \frac{d\bar{\mu}}{\bar{\mu}}\gamma_q(\bar{\mu}), \notag\\
    S_K(t)=&s\left(\frac{M_B}{\sqrt{2}}x_2,b_2\right)
    +s\left(\frac{M_B}{\sqrt{2}}(1-x_2),b_2\right) \notag\\
    &+2\int_{1/b_2}^t\frac{d\bar{\mu}}{\bar{\mu}}\gamma_q(\bar{\mu}),
\end{align}
where $\gamma_q=-\alpha_s/\pi$ is the quark anomalous dimension, and the expression for $s(Q,b)$ can be written as~\cite{Li:1994iu}
\begin{align}
    s(Q,b)=&\frac{A^{(1)}}{2\beta_1}\hat{q}\ln\left(\frac{\hat{q}}{\hat{b}}\right)
    +\frac{A^{(2)}}{4\beta_1^2}\left(\frac{\hat{q}}{\hat{b}}-1\right)
    -\frac{A^{(1)}}{2\beta_1}\left(\hat{q}-\hat{b}\right) \notag\\
    &-\frac{A^{(1)}\beta_2}{4\beta_1^3}\hat{q}\left[\frac{\ln(2\hat{b})+1}{\hat{b}}
    -\frac{\ln(2\hat{q})+1}{\hat{q}}\right] \notag\\
    &-\left[\frac{A^{(2)}}{4\beta_1^2}-\frac{A^{(1)}}{4\beta_1}
    \ln\left(\frac{e^{2\gamma-1}}{2}\right)\right]
    \ln\left(\frac{\hat{q}}{\hat{b}}\right) \notag\\
    &+\frac{A^{(1)}\beta_2}{8\beta_1^3}\left[\ln^2(2\hat{q})-\ln^2(2\hat{b})\right],
\end{align} 
where
\begin{align}
    \hat{q}&=\ln\left[Q/(\sqrt{2}\Lambda)\right], \quad 
    \hat{b}=\ln\left(1/(b\Lambda)\right), \notag\\
    \beta_{1}&=\frac{33-2n_f}{12}, \quad \beta_{2}=\frac{153-19n_f}{24}, \notag\\
    A^{(1)}&=\frac{4}{3}, \quad A^{(2)}=\frac{67}{9}-\frac{\pi^{2}}{3}
    -\frac{10n_f}{27}+\frac{8}{3}\beta_1\ln\left(\frac{e^{\gamma}}{2}\right).
\end{align}

Threshold resummation factor~\cite{Kurimoto:2001zj}
\begin{align}
    S_t(x)=\frac{2^{1+2c}\Gamma(3/2+c)}{\sqrt{\pi}\Gamma(1+c)}[x(1-x)]^c,
\end{align}
with $c=0.3$.

NLO hard kernel correction factor at twist-two~\cite{Li:2012nk}
\begin{align}
    &F_{\mathrm{T}2}^{(1)}\left(x_i,\mu,t,q^2\right)=\frac{\alpha_s(t)C_F}{4\pi} 
    \Bigg[\frac{21}{4}\ln\frac{\mu^2}{M_B^2} \notag\\
    &-\left(\ln\frac{M_B^2}{\xi_1^2}+\frac{13}{2}\right)\ln\frac{t^2}{M_B^2} 
    +\frac{7}{16}\ln^2(x_1x_2)+\frac{1}{8}\ln^2x_1 \notag\\
    &+\frac{1}{4}\ln x_1\ln x_2 +\left(2\ln\frac{M_B^2}{\xi_1^2}
    +\frac{7}{8}\ln\eta-\frac{1}{4}\right)\ln x_1 \notag\\
    &+\left(\frac{7}{8}\ln\eta -\frac{3}{2}\right)\ln x_2 
    +\left(\frac{15}{4}-\frac{7}{16}\ln\eta\right)\ln\eta \notag\\
    &-\frac{1}{2}\ln\frac{M_B^2}{\xi_1^2}\left(3\ln\frac{M_B^2}{\xi_1^2}+2\right) 
    +\frac{101\pi^2}{48}+\frac{219}{16}\Bigg],
\end{align}
with
\begin{align}
    \mu=\left\{\exp\left[c_1+\left(\ln\frac{M_B^2}{\xi_1^2}+\frac{5}{4}\right)
    \ln\frac{t^2}{M_B^2}\right]x_1^{c_2} x_2^{c_3}\right\}^{2/21} \cdot t,
\end{align}
with
\begin{align}
    c_1=&-\left(\frac{15}{4}-\frac{7}{16}\ln\eta\right)\ln\eta+\frac{1}{2}
    \ln\frac{M_B^2}{\xi_1^2}\left(3\ln\frac{M_B^2}{\xi_1^2}+2\right) \notag\\
    &-\frac{101\pi^2}{48}-\frac{219}{16}, \notag\\
    c_2=&-\left(2\ln\frac{M_B^2}{\xi_1^2}+\frac{7}{8}\ln\eta-\frac{1}{4}\right), \notag\\
    c_3=&-\frac{7}{8}\ln\eta+\frac{3}{2}.
\end{align}

NLO hard kernel correction factor at twist-three~\cite{Cheng:2014fwa}
\begin{align}
    &F_{\mathrm{T}3}^{(1)}\left(x_i,\mu,t,q^2\right)=\frac{\alpha_s(t)C_F}{4\pi} 
    \Bigg[\frac{21}{4}\ln\frac{\mu^2}{M_B^2} \notag\\
    &-\frac{1}{2}(6+\ln\frac{M_B^2}{\xi_1^2})\ln\frac{t^2}{M_B^2} 
    +\frac{7}{16}\ln^2 x_1-\frac{3}{8}\ln^2 x_{2}\notag\\
    &+\frac{9}{8}\ln x_1 \ln x_2+\left(\ln\frac{M_B^2}{\xi_1^2}
    +\frac{15}{8}\ln\eta-\frac{29}{8}\right)\ln x_1 \notag\\
    &+\left(\ln \frac{M_B^2}{\xi_2^2}+\frac{9}{8}\ln\eta-\frac{25}{16}\right)\ln x_2 
    +\frac{1}{2}\ln \frac{M_B^2}{\xi_1^2}\notag\\
    &-\frac{1}{4} \ln^2 \frac{M_B^2}{\xi_1^2}
    +\ln \frac{M_B^2}{\xi_2^2}-\frac{9}{8}\ln\eta-\frac{1}{8}\ln^2\eta
    +\frac{37\pi^2}{32}+\frac{91}{32}\Bigg], 
\end{align}
with
\begin{align}
    \mu=\left\{\exp\left[c_1+\left(\frac{1}{2}\ln \frac{M_B^2}{\xi_1^2}-\frac{9}{4}\right)
    \ln\frac{t^2}{M_B^2}\right]x_1^{c_2} x_2^{c_3}\right\}^{2/21} \cdot t,
\end{align}
with
\begin{align}
    c_1=&-\left(\frac{1}{2}-\frac{1}{4}\ln \frac{M_B^2}{\xi_1^2}\right)\ln \frac{M_B^2}{\xi_1^2}
    +\left(\frac{9}{8}+\frac{1}{8}\ln\eta\right)\ln\eta \notag\\
    &-\frac{379}{32}-\frac{167\pi^2}{96},\notag\\
    c_2=&\frac{29}{8}-\ln \frac{M_B^2}{\xi_1^2}-\frac{15}{8}\ln\eta,\notag\\
    c_3=&\frac{25}{16}-\frac{9}{8}\ln\eta, 
\end{align}
where $\xi_1=25M_B$ and $\xi_2^2=M_B^2$.


\end{document}